\documentclass[twocolumn,showpacs,preprintnumbers,amsmath,amssymb]{revtex4}
\usepackage{graphicx}
\usepackage{dcolumn}
\usepackage{bm}
\usepackage[latin1]{inputenc}
\usepackage{color}
\usepackage{subfigure}
\newcommand{\comment}[1]{}

\begin{document}
\setlength{\unitlength}{0.7\textwidth} \preprint{}

\title{Angular multiscale statistics of Lagrangian trajectories in turbulence}

\author{Wouter J.T. Bos{$^1$}, Benjamin Kadoch{$^2$} and Kai Schneider{$^3$}}

\affiliation{$^1$ LMFA,  CNRS UMR 5509, Ecole Centrale de Lyon, Universit\'e de Lyon, Ecully, France}
\affiliation{$^2$ IUSTI, CNRS UMR 7343, Aix-Marseille University, Marseille, France}
\affiliation{$^3$ M2P2,  CNRS UMR 7340 \& CMI, Aix-Marseille University, Marseille, France}

\begin{abstract}
The angle between subsequent particle displacement increments is evaluated as a function of the timelag in isotropic turbulence. It is shown that the evolution of this angle contains two well-defined power-laws, reflecting the multi-scale dynamics of high-Reynolds number turbulence. The probability density function of the directional change is shown to be self-similar and well approximated by an analytically derived model assuming Gaussianity and independence of the velocity and the Lagrangian acceleration.  
\end{abstract}



\pacs{47.27.Jv, 47.27.Gs}
\maketitle

Advances in experimental devices and numerical simulations over the last two decades have opened the way to a Lagrangian characterization of turbulent flows \cite{Yeung2002,Mordant2004,Toschi2009}. The structural description of the statistical dynamics of turbulence has thereby shifted from the investigation of spatial correlations of instantaneous velocity fields to the study of temporal correlations along fluid particle trajectories. In the Lagrangian reference frame, the spatio-temporal complexity of turbulence manifests itself through the spiraling chaotic motion of fluid particles, changing direction at every timescale. This directional change of Lagrangian tracers, as a function of the timelag between two observations, is the subject of the present work. Instantaneous measures of the curvature in turbulence have been investigated in the past five years for academic turbulent flows, both in three \cite{Braun2006,Xu2007} and in two space dimensions \cite{Wilczek2008,Kadoch2011-2}. Curvature is dominated 
by the small-scale 
structures 
and contains 
only little information on the multiscale dynamics of turbulent flows. Multi-scale dynamics can be measured by Lagrangian structure functions \cite{Yeung2002,Toschi2009}, but those do not contain any direct information on the curvature of the trajectories. 

The related measure which represents the coarse grained curvature over a time 
interval was only recently introduced by Burov {\it et al.} \cite{Burov2013}. More precisely, in this last work the directional change of a particle was introduced, and the characteristics of this new measure were investigated in various types of random walks. In the present work, we will show how this measure can characterize the time-correlation of the direction of a fluid particle in a turbulent flow. In particular will we show how the multi-scale character of a turbulent flow can be revealed by considering the timelag dependence of the directional change.

\begin{figure}
\centering
\setlength{\unitlength}{.5\textwidth}
\includegraphics[width=1.\unitlength,angle=0]{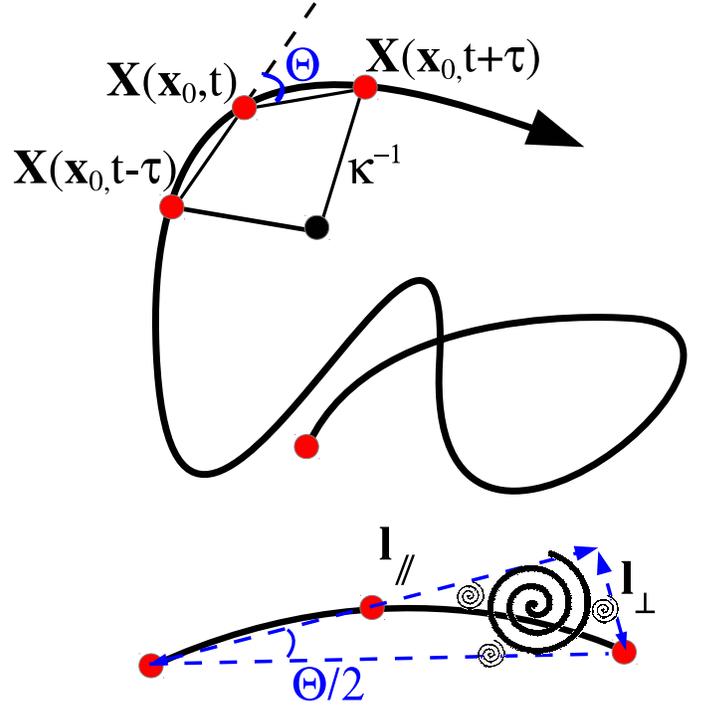}
\caption{Top: definition of the angle between subsequent Lagrangian particle increments. Bottom: short time evolution and definition of the lengthscales $l_\perp$ and $l_\parallel$. 
\label{fig:sketch} }     
\end{figure}

We define the Lagrangian spatial increment as
\begin{equation}
 \delta \bm X(\bm x_0,t,\tau)=\bm X(\bm x_0,t)-\bm X(\bm x_0,t-\tau)
\end{equation}
where $\bm X(\bm x_0,t)$ is the position of a fluid particle at time $t$, passing through point $\bm x_0$ at the reference time $t=t_0$ and advected by a velocity field $\bf u$, i.e. $d{\bm X}/dt = {\bf u}$ . The cosine of the angle $\Theta(t,\tau)$ between subsequent particle increments, introduced in \cite{Burov2013}, is 
\begin{equation}\label{eq:cos}
 \cos(\Theta(t,\tau))=\frac{\delta \bm X(\bm x_0,t,\tau)\cdot\delta \bm X(\bm x_0,t+\tau,\tau)}{|\delta \bm X(\bm x_0,t,\tau)|~|\delta \bm X(\bm x_0,t+\tau,\tau)|}.
\end{equation}
The angle is illustrated in Figure \ref{fig:sketch} (top). Rather than considering its instantaneous evolution, its averaged absolute value is of particular interest in an isotropic random velocity field. The ensemble average will be denoted in the following by
\begin{equation}
\theta(\tau)\equiv\left<|\Theta(t,\tau)|\right>.
\end{equation}
We omitted the time-dependence since we will consider statistically stationary flow in the following. For short time lags, $\theta(\tau)$ should be close to zero, whereas for times long compared to the correlation time associated with the spiraling motion $\theta(\tau)$ should tend to $\pi/2$ by symmetry. 

For short times the instantaneous angle $\Theta(\tau,t)$ is related to the curvature $\kappa$  (see Figure \ref{fig:sketch} (top)) by the relation
\begin{equation}
\kappa(t)=\lim_{\tau\rightarrow 0}  \frac{|\Theta(t,\tau)|}{2\tau\|\bm u(t)\|},
\end{equation}
with $\bm u$ being the velocity. How the angle varies in between the short and long-time limits is the main subject of the present work and we will show that the dependence of $\theta(\tau)$  on the timelag contains the signature of the multi-scale dynamics of a turbulent flow. 

\begin{figure}
\centering
\setlength{\unitlength}{.5\textwidth}
\includegraphics[width=1.\unitlength,angle=0]{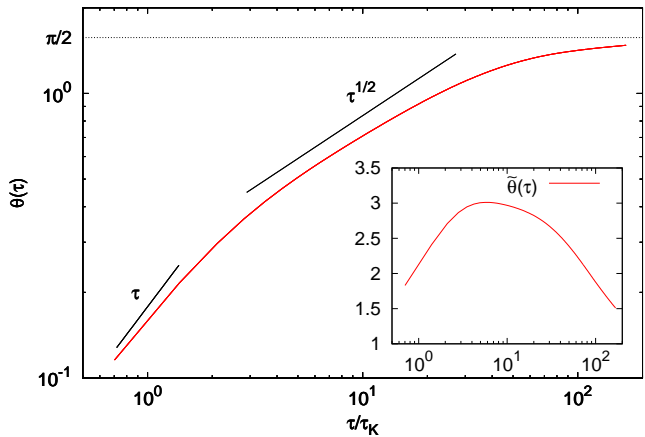}
\caption{The average angle $\theta$ as a function of the timelag $\tau$ normalized by the Kolmogorov time-scale $\tau_K$. In the inset the compensated angle, $\tilde \theta(\tau)\equiv \theta(\tau)\sigma_u/(\epsilon \tau)^{1/2}$ is plotted.
\label{fig:Fig2} }     
\end{figure}

The database used to investigate the behaviour of $\theta(\tau)$ is described in \cite{Kamps2011,Wilczek2013}. The simulation was carried out using standard pseudo-spectral techniques, following $8.10^6$ fluid particles in a statistically stationary isotropic turbulent flow during $5.8$ integral timescales in a periodic cube of dimension $2\pi$. The resolution is $1024^3$ gridpoints. The integral timescale is $2.1$ and the Kolmogorov timescale $\tau_K=(\nu/\epsilon)^{1/2}= 0.036$, where $\epsilon=0.31$ is the mean dissipation rate and $\nu=4.10^{-4}$ the kinematic viscosity. The Lagrangian integral timescale is of the order of the Eulerian integral timescale. The Taylor-scale Reynolds number is $R_\lambda=225$. 

Figure \ref{fig:Fig2} shows $\theta(\tau)$ in double-logarithmic representation. The angle increases monotonuously from zero to $\pi/2$, and this latter value is approached for values of $\tau$ of the order of the Lagrangian integral timescale. Two power-laws can be identified in this graph, with a cross-over around twice the Kolmogorov timescale. The origin of these power-laws will now be elucidated.

For our phenomenological explanation, we consider high-Reynolds-number isotropic turbulence, containing flow structures on a wide range of different scales. We consider short timelags $\tau\ll T$, where $T$ is the Lagrangian integral timescale of the flow.  In this limit, the angle $\Theta(t,\tau)$ can be approximated using a Taylor-expansion by,
\begin{equation}
\frac{l_\perp}{l_\parallel} \approx |\tan(\Theta/2)|\approx |\Theta/2|.
\end{equation}
where $l_\perp$ and $l_\parallel$ are shown in Fig. \ref{fig:sketch}, (bottom), and correspond to the absolute values of the distance travelled parallel with, and perpendicular to, the initial displacement increment, respectively, over a time interval $2\tau$. The values of $l_\parallel$ and $l_\perp$ can be estimated, again using a Taylor expansion, to be 
\begin{equation}
 l_\parallel\approx 2 U(t,\tau) \tau~~~~~~  l_\perp\approx 2\tau^2 a_\perp(t,\tau),
\end{equation}
with $U(t,\tau)$ and $a_\perp(t,\tau)$ the absolute values of the velocity and the acceleration perpendicular to the velocity, respectively, coarse-grained over a time $\tau$ along the fluid particle trajectory.  Without loss of generality, we will write $U(t,\tau)$ and $a_\perp(t,\tau)$ as
\begin{equation}
 U(t,\tau)= \sigma_u(\tau)\xi_u(t,\tau)~~~~~~   a_\perp(t,\tau)= \sigma_a(\tau)\xi_a(t,\tau),
\end{equation}
where $\sigma_u^2(\tau)$ and $\sigma_a^2(\tau)$ are the variance of the coarse-grained velocity and perpendicular acceleration, respectively. The quantities $\xi_u(t,\tau)$ and $\xi_a(t,\tau)$ are positive random variables with unit mean value and unit mean variance. We thereby obtain,
\begin{equation}\label{eq:thetaabs}
|\Theta(t,\tau)|\approx 2\tau \frac{\sigma_a(\tau)\xi_a(t,\tau)}{\sigma_u(\tau)\xi_u(t,\tau)}. 
\end{equation}
We assume the velocity and the acceleration independent, a reasonable assumption at very high Reynolds numbers, as long as $\tau\ll T$. Without coarse-graining this assumption was also used in reference \cite{Xu2007} to model the curvature in isotropic turbulence. Using this assumption we find
\begin{equation}\label{eq:thetaabs2}
\theta(\tau)\approx 2\tau \frac{\sigma_a(\tau)}{\sigma_u(\tau)}. 
\end{equation}
Since for  $\tau\ll T$ the velocity is roughly constant over the time-interval, $\sigma_u(\tau)\approx U_{\textrm{rms}}$. However, $a_\perp$ is dominantly determined by the small scales and fluctuates rapidly. Only for $\tau$ small with respect to the smallest Lagrangian time-scale, the Kolmogorov scale, can we consider $\sigma_a(\tau)\approx (a_\perp)_{\textrm{rms}}$, i.e., independent of $\tau$. For these very short time-scales, we have thus
\begin{equation}\label{eq:thetaabs3}
\theta(\tau)\approx 2\tau \frac{\sigma_a}{\sigma_u} ~~~\textrm{for}~\tau\ll\tau_K.
\end{equation}
where $\sigma_a$ and $\sigma_u$ are the total rms perpendicular acceleration and velocity, respectively.  The linear relation between $\theta(\tau)$ and $\tau$ is well observed in Figure \ref{fig:Fig2}. We can further express this in terms of quantities which are easy to determine experimentally. Assuming classical scaling \cite{Heisenberg}, the acceleration variance is given by the relation
\begin{equation}\label{eq:aepsnu}
 \sigma^2_a\sim \frac{\epsilon^{3/2}}{\nu^{1/2}},
\end{equation}
a relation which can be refined to take into account intermittency corrections \cite{Vedula1999,Ishihara2007,Gotoh1999}. Omitting these corrections, expressions (\ref{eq:thetaabs3}) and (\ref{eq:aepsnu}) yield,
\begin{equation}
\theta(\tau)\sim \frac{\tau}{T}R_\lambda^{1/2} ~~~\textrm{for}~\tau\ll\tau_K.
\end{equation}

At timescales larger than $\tau_K$, but smaller than $T$, i.e., in the inertial interval, the above approximations to obtain (\ref{eq:thetaabs2}) are still valid. However, the subsequent approximation, that  $\sigma_a(\tau)$ is independent of $\tau$ is not valid anymore. Indeed, the perpendicular acceleration fluctuates rapidly in time, on a time-scale of the order of $\tau_K$. Coarse-graining the acceleration over an interval $\tau>\tau_K$, the influence of the more rapidly fluctuating scales is filtered out. Indeed, even if their contribution to the rms acceleration is dominant, if the coarse-graining is performed before considering the norm, positive and negative contributions will cancel each other. The remaining variance will be predominantly caused by scales with a time-scale larger than,  or comparable to $\tau$. Following classical Kolmogorov phenomenology \cite{Kolmogorov,Tennekes}, the acceleration induced by inertial range structures with typical timescale $\tau$ will be of the order 
\begin{equation}\label{eq:aofl}
 \sigma_a(\tau)\sim(\epsilon/\tau)^{1/2}.
\end{equation}
This estimate is obtained by neglecting the viscous contribution to the acceleration, a reasonable assumption even near the dissipation range scales \cite{Ishihara2007}, and realizing that the acceleration is due to pressure forces, which satisfy, at inertial range scales to a good approximation Kolmogorov-scaling \cite{Gotoh2001,Monin}.
 The scale of such eddies is proportional to
\begin{equation}
 l(\tau)\sim \tau^{3/2}\epsilon^{1/2}.
\end{equation}
The reciprocal dependence of the acceleration variance on $\tau$ in expression (\ref{eq:aofl}) illustrates that the smallest scales are most efficient in accelerating the fluid particles. After the influence of the scales smaller than $\tau$ is removed by the coarse graining, the remaining dominant contribution is caused by the smallest scales still present, i.e., with timescale $\tau$. It is therefore those scales, with correlation-time $\tau$ which will deviate particles from their trajectory over a lengthscale of the order of the correlation-length of the structures. This phenomenological picture is illustrated in Fig. \ref{fig:sketch}, bottom, where it can be understood intuitively that scales of the size $l\ll l(\tau)$ are too small to efficiently contribute to a perpendicular displacement averaged over a time-interval $\tau$. 

Combining (\ref{eq:aofl}) and (\ref{eq:thetaabs}) we obtain in the inertial range
\begin{equation}\label{eq:thetaIR}
\theta(\tau)\sim \tau^{1/2} \frac{\epsilon^{1/2}}{\sigma_u}\sim \left(\frac{\tau}{T}\right)^{1/2}   ~~~\textrm{for}~\tau_K\ll\tau \ll T.
\end{equation}
Again, this scaling is observable in Figure \ref{fig:Fig2}, even though the power-law is less well present than in the dissipation range. This is better appreciated by considering the compensated angle, $\tilde \theta(\tau)\equiv \theta(\tau)\sigma_u/(\epsilon \tau)^{1/2}$, plotted in the inset of the figure.  The slow emergence of inertial ranges with the Reynolds number in Lagrangian statistics is fairly common \cite{Yeung2006} and is was recently even argued that they might be non-existent \cite{Falkovich2012}. In the present case, the emergence of a plateau is undeniable. This might be because the inertial range scaling of the mean-angle $\theta(\tau)$ is not directly related to the Lagrangian structure functions. Indeed, the scaling is induced by considering the coarse-grained Lagrangian acceleration, a quantity of which the scaling is related to that of the Eulerian pressure gradient.

\begin{figure}
\centering
\setlength{\unitlength}{.45\textwidth}
\subfigure[]{\includegraphics[width=1.\unitlength,angle=0]{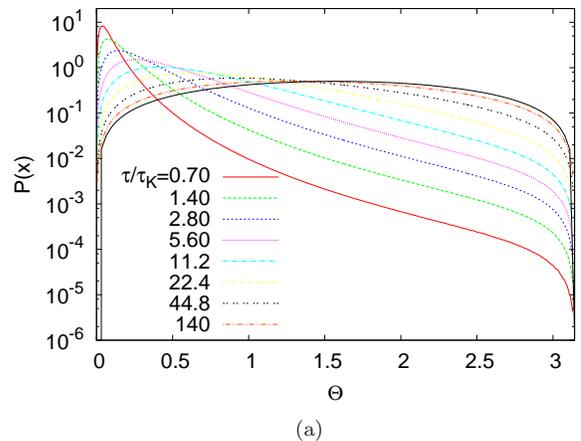}}
\subfigure[]{\includegraphics[width=1.\unitlength,angle=0]{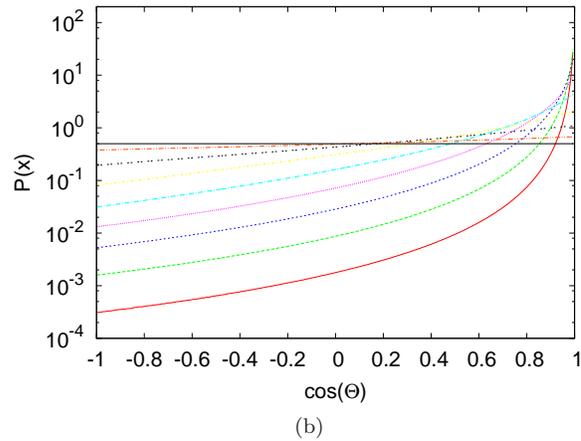}}
\caption{PDFs of (a) $\Theta$ and (b) $\cos(\Theta)$ for different timelags. Solid black lines indicate the long time asymptotic form of the PDFs.
\label{fig:3} }     
\end{figure}

\begin{figure}
\centering
\setlength{\unitlength}{.45\textwidth}
\subfigure[]{\includegraphics[width=1.\unitlength,angle=0]{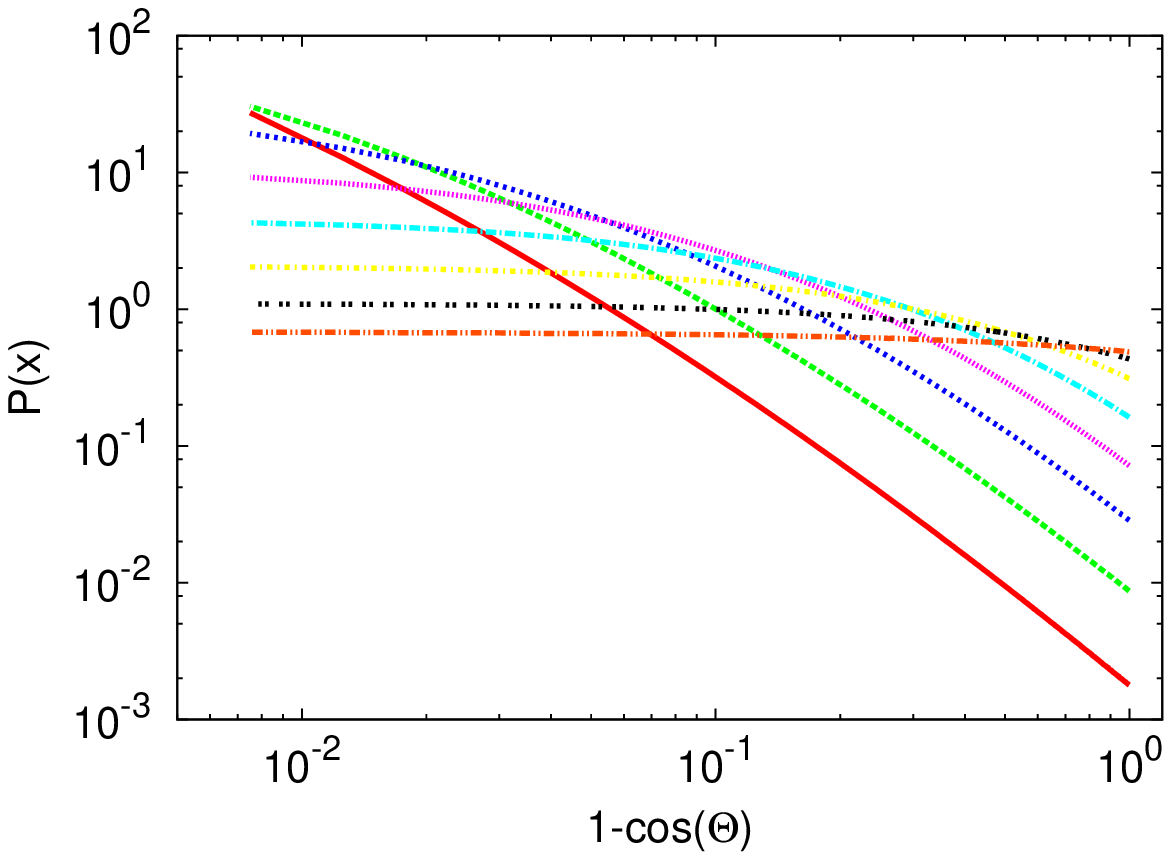}}
\subfigure[]{\includegraphics[width=1.\unitlength,angle=0]{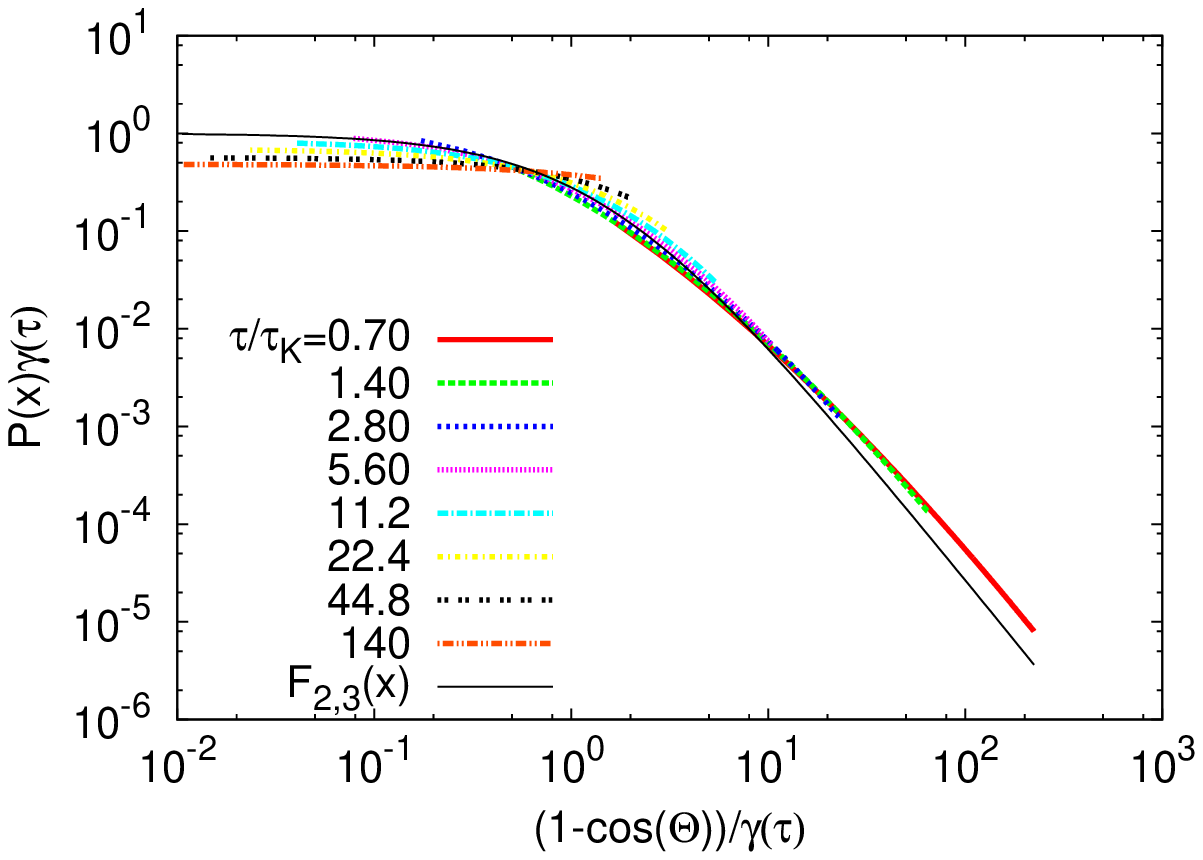}}
\caption{(a) PDFs of $1-\cos(\Theta)$. (b) The same PDFs normalized and compared to the analytical prediction.
\label{fig:4} }     
\end{figure}

The above arguments and results considered the average value $\theta(\tau)$ only. Further information, in particular on higher order moments, is contained in the probability density function (PDF) of the instantaneous angle and its evolution with $\tau$. Those PDFs of the angle $\Theta(t,\tau)$ and its cosine are shown in Figure  \ref{fig:3}. It is observed that the $P_\tau(\Theta)$ for small $\tau$ consists of a peak near zero, whereas for long-times a symmetric distribution between $0$ and $\pi$ is obtained. This latter distribution corresponds to the distribution between two randomly chosen vectors in three-dimensions. Its distribution is given by $P_\infty(\Theta)= \sin(\Theta)/2$. The distribution of the cosine of the angle between two random vectors is thus equidistributed so that $P_\infty(\cos(\Theta))=1/2$. It is observed that these two long-time distributions are approached for long times in Figure \ref{fig:3}. We will now show how we can predict the short-time, small $\Theta(\tau,t)$ behaviour of 
the 
PDFs. In particular will we consider the PDF of $1-\cos(\Theta)$. Since the cosine of small deviations in $\Theta$ gives values near unity, $1-\cos(\Theta)$ directly measures the magnitude of the 
directional change. In addition, it is easy to compare the PDF of this quantity to the long-time limit consisting of a straight line. The PDFs of $1-\cos(\Theta)$ are shown in Figure \ref{fig:4}(a) in double logarithmic representation. To explain their shape and their evolution with $\tau$, we use a Taylor expansion for small $\Theta$, 
\begin{equation}
1-\cos(\Theta(t,\tau))\approx \frac{1}{2} \Theta(t,\tau)^2\approx  2\tau^2 \frac{\sigma^2_a(\tau)\xi^2_a(t,\tau)}{\sigma^2_u(\tau)\xi^2_u(t,\tau)},
\end{equation}
where we used expression (\ref{eq:thetaabs}). If we assume $\bm a_\perp$ to satisfy a Gaussian distribution, which is only a good approximation for the core of the PDF,  and if we further assume $\bm u$ to be uncorrelated with $\bm a_\perp$, and also multivariate Gaussian \cite{Xu2007}, then both $\xi_a^2$ and $\xi_u^2$ satisfy a $\chi$-squared distribution. For a given velocity vector in $3D$, having 3 degrees of freedom, the perpendicular acceleration is confined to the plane perpendicular to the velocity and is a 2-component vector. The ratio of two properly normalized independent $\chi^2$-distributed quantities with $n,m$ degrees of freedom, respectively, is given by an $F_{n,m}$ Fischer-distribution. We expect $1-\cos(\Theta)$ therefore to be given by an $F_{2,3}$ distribution. More precisely,
\begin{equation}
\gamma(\tau)P_{1-\cos(\Theta(t,\tau))}(x/\gamma(\tau))= F_{2,3}(x),
\end{equation}
where $\gamma_\tau=\theta(\tau)^2/3$ and $\theta(\tau)$ is shown in Figure \ref{fig:Fig2}. It is observed in Figure \ref{fig:4}(b) that the agreement with the prediction is fairly satisfactory considering the assumptions we made in the derivation of the shape of the PDF. Note that no adjustable parameters were used to fit the PDF to the $F$-distribution.

The results obtained in the present investigation show that the time-series of the Lagrangian position can reveal the inertial range structure of turbulence through the timelag-dependence of the quantity $\theta(\tau)$. In particular do we show how Kolmogorov's inertial range theory is linked to the angular statistics of Lagrangian fluid particle trajectories.

The present framework will allow experimentalists to verify the scaling of Lagrangian statistics in very-high-Reynolds numbers flows, even if the measurement techniques are not sufficiently rapid to resolve down to the Kolmogorov scale. Indeed, no measurements of the instantaneous velocity or acceleration are needed, only Lagrangian position measurements sufficiently sampled to resolve the intertial range timescales. A further come-out of this investigation are the scale-dependent measures for the mean-angle and the probability density functions, which will allow to more accurately model the topology of Lagrangian trajectories in dispersion models.  

The measure we investigated in the foregoing allows a different angle of attack on the simultaneous characterisation of the multi-scale character of turbulence and the scale dependent curvature of Lagrangian fluid particle trajectories. In this light an interesting perspective is to clarify the link between the current work and the results obtained using the recently introduced longitudinal and transversal Lagrangian structure functions \cite{Leveque2014}.

\emph{Acknowledgments.} The authors are indebted to Oliver Kamps and Michael Wilczek who provided us the DNS data used in the present investigation.


\end{document}